\documentclass[onecolumn, draftcls, 12pt]{IEEEtran}

\usepackage{amsmath}
\usepackage{cases}
\usepackage{amsfonts}
\usepackage{amssymb}
\usepackage{boxedminipage}
\usepackage{graphicx}
\usepackage[usenames]{color}
\usepackage{array,color}
\usepackage{colortbl}
\usepackage{extarrows}
\usepackage{bm}
\usepackage{amsmath}
\begin{document}

\title{\Large New inequalities of Mill's ratio  and Application  to The Inverse Q-function Approximation}

\author{ Pingyi~Fan, \IEEEmembership{Senior Member,~IEEE}\\
fpy@tsinghua.edu.cn,\\
Department of Electronic Engineering, Tsinghua University, Beijing,  China;\\
}


\maketitle

\begin{abstract}

In this paper, we investigate the Mill's ratio estimation problem and get two new inequalities. Compared to the well known results obtained by Gordon, they becomes  tighter. Furthermore, we also discuss the inverse Q-function approximation problem and present some useful results on the inverse solution. Numerical results confirm the validness of our theoretical analysis. In addition, we also present a conjecture on the bounds of inverse solution on Q-function.
\end{abstract}

\section{Introduction}

The Gaussian Q-function is always used to present the probability that a standard normal random variable exceeds a positive value  $x$ and
is defined by
\begin{equation}
Q(x)=\frac{1}{\sqrt{2\pi}}\int_x^{\infty}e^{-u^2/2}du
\end{equation}
Since the prevalence of normal random variables, the Q-function, as one of the most important integrals, is usually
encountered in applied mathematics, statistics, and engineering. However, it is very difficult to handle
mathematically due to its non-elementary integral form which cannot be expressed as a finite composition of simple
functions.  For this reason, a lot of works have been on the development of
approximations and bounds for the Q-function. The well known approximation form was first given by Gordon \cite{Gordon1941}, usually referred to "Mills ratio inequalities". Later on, Birnhaum improved Gordon's lower bound \cite{Birnbaum1942} and Sampford improved Gordon's upper bound \cite{samford1953}.  Baricz \cite{Baricz2007} presented  new proofs on Birnhaum and Sampford's results by using monotonicity properties of some functions involving the Mill's ratio of standard normal law. In \cite{sundberg1979}, Borjesson and Sundberg extended the results of Birnhaum and Sampford by computer search to find some explicit approximation functions to Q-function.  The same parameter selection problem was treated by Boyd \cite{boyd1959}.  Tate \cite{Tate1953} also presented some inequalities for real positive number and negative number. Some works focused on using a sum of multiple terms to approximate the Q-function \cite{shenton1954}\cite{ray1963}\cite{alzer1997}\cite{pinelis2002a}\cite{pinelis2002b}\cite{dumbgen2010}.  Some works derived the  Chernoff-type bounds of the Q-function, including upper and lower bounds \cite{chang2011}\cite{cote2012}.
In this paper, we will focus on the improvement of Mills' ratio inequalities by modifying the multiplying factor function of $e^{-x^2/2}$ while keeping the type of original form of Mills' inequalities.  We get two improved inequalities, including one upper bound and one lower bound. Compared to the well known inequalities, the new developed lower bound becomes much tighter when integral variable $x$ is relatively large.  In addition, we also consider the approximation of the inverse solution of Q-function and obtain some useful results, among them  one setting up a close relationship between the information entropy and Q-function.

\textbf{\textit{Theorem} 1} (Mills' Ratio inequality\cite{Gordon1941}\cite{Fan-SCT})\\

For arbitrary positive number $x>0$, the inequalities
\begin{equation}
\frac{x}{1+x^2}e^{-x^2/2} < \int_x^{\infty}e^{-u^2/2}du< \frac{1}{x}e^{-x^2/2}
\end{equation}
are valid. In particular,
\begin{equation}
\int_x^{\infty}e^{-u^2/2}du\approx\frac{1}{x}e^{-x^2/2}
\end{equation}
holds when $ x\rightarrow \infty $.

\textbf{\textit{Theorem} 2}( Birnbaum and Sampford)\\
The inequalities

\begin{equation}
\frac{2}{\sqrt{x^2+4}+x}e^{-x^2/2}<\int_x^{\infty}e^{-u^2/2}du <\frac{4}{\sqrt{x^2+8}+3x}e^{-x^2/2}
\end{equation}
holds for all $x>0$.

\textbf{\textit{Theorem} 3} (New Mills' Ratio inequality)\\

The inequalities
\begin{equation} \label{new upp}
\int_x^{\infty}e^{-u^2/2}du<\frac{1}{\sqrt{1+x^2}}e^{-x^2/2}
\end{equation}
is valid for all $x>\sqrt{\frac{\sqrt{5}-1}{2}}$  and
\begin{equation} \label{new low}
\frac{1+x^2}{x(2+x^2)}e^{-x^2/2}<\int_x^{\infty}e^{-u^2/2}du
\end{equation}
is valid for all $x>\sqrt{2}$.

In particular,
\begin{equation}
\int_x^{\infty}e^{-u^2/2}du\approx\frac{1}{\sqrt{1+x^2}}e^{-x^2/2}
\end{equation}
holds when $x\rightarrow \infty$.

In fact, the upper bound in Theorem 3 is worse than that given in Theorem 2, but we still like to keep it since it has a relatively simple expression and is also useful in the estimation of the inverse Q-function, which will be shown in Section IV.

By combing the results of Theorem 2 and Theorem 3, we have

\textbf{\textit{Corollary} 1}.\\

The inequalities
\begin{equation}
f_1(x) < \int_x^{\infty}e^{-u^2/2}du< f_2(x)
\end{equation}
are valid for all $x> 0$, where $f_1(x)$ and $f_2(x)$ are given as follows
\begin{equation}
f_1(x)=\left\{\begin{aligned}
         &\frac{2}{\sqrt{x^2+4}+x}~~~\textrm{if}~0<x\leq \sqrt{2} \\
         &\frac{1+x^2}{x(2+x^2)}~~~\textrm{if}~x>\sqrt{2}
                 \end{aligned} \right.
\end{equation}
 and
\begin{equation}
f_2(x)=\frac{4}{\sqrt{x^2+8}+3x}e^{-x^2/2}
\end{equation}
In particular,
\begin{equation}
\int_x^{\infty}e^{-u^2/2}du\approx\frac{1}{\sqrt{1+x^2}}e^{-x^2/2}
\end{equation}
holds when $x\rightarrow \infty$.

\section{Proof of the main result}
\label{sec:proof}

\subsection{Proof of Theorem 3}
Let us define a function
\begin{equation}
g(u)=-\frac{1}{\sqrt{1+u^2}}
\end{equation}
for all $u>0$.

Differentiation yields
\begin{equation}
\frac{dg(u)}{du}=\frac{u}{(1+u^2)^{3/2}} \nonumber
\end{equation}

Thus, we have
\begin{eqnarray}
\int_x^{\infty}\frac{u}{(1+u^2)^{3/2}}e^{-u^2/2}du = \int_x^{\infty} e^{-u^2/2} dg(u) \nonumber \\
=g(u)e^{-u^2/2}|_x^{\infty}-\int_x^{\infty}g(u)e^{-u^2/2}u du \nonumber \\
=\frac{1}{\sqrt{1+x^2}}e^{-x^2/2}-\int_x^{\infty}\frac{u}{\sqrt{1+u^2}}e^{-u^2/2} du
\end{eqnarray}
By reorganizing the integral equality above, we get
\begin{equation}
\int_x^{\infty} [\frac{u}{(1+u^2)^{3/2}}+\frac{u}{(1+u^2)^{1/2}}]e^{-u^2/2} du =\frac{1}{\sqrt{1+x^2}}e^{-x^2/2}
\end{equation}
That is,
\begin{equation}\label{equality 1}
\int_x^{\infty}\frac{u(2+u^2)}{(1+u^2)^{3/2}}e^{-u^2/2} du =\frac{1}{\sqrt{1+x^2}}e^{-x^2/2}
\end{equation}

It is easy to find that if $u>\sqrt{\frac{\sqrt{5}-1}{2}}$,
\begin{equation}
\frac{u(2+u^2)}{(1+u^2)^{3/2}}>1
\end{equation}
In fact, by defining $g_1(u)=u(2+u^2)$, $g_2(u)=(1+u^2)^{3/2}$ and $g_3(u)=\frac{g_1(u)}{g_2(u)}$ for all $u>0$, we have
\begin{eqnarray}
g_1^2(u)-g_2^2(u)&=&u^6+4u^4+4u^2-(u^6+3u^4+3u^2+1)\nonumber \\
&=&u^4+u^2-1>0
\end{eqnarray}
if $u>\sqrt{\frac{\sqrt{5}-1}{2}}\approx 0.7862$.

By using the results above, Eqn. (\ref{equality 1}) becomes the following inequality
\begin{equation}
\int_x^{\infty}e^{-u^2/2} du <\frac{1}{\sqrt{1+x^2}}e^{-x^2/2},
\end{equation}
which is valid for $x>\sqrt{\frac{\sqrt{5}-1}{2}}\approx 0.7862$. Therefore,  the first inequality Eqn. (\ref{new upp}) is proved.

On the other hand, it is not hard to get $g_3(u)$ is monotonically decreasing for $u\geq\sqrt{2}$.\\

  Since
\begin{eqnarray}
\frac{dg_3(u)}{du}&=&\frac{(2+3u^2)(1+u^2)^{3/2}-3u^2(2+u^2)(1+u^2)^{1/2}}{(1+u^2)^3}\nonumber \\
&=&\frac{2-u^2}{(1+u^2)^{5/2}}
\end{eqnarray}
If $u\geq \sqrt{2}$, then $\frac{dg_3(u)}{du}<0$, resulting in  that $g_3(u)$ is monotonically decreasing for $u\geq\sqrt{2}$.

In this case, we have

\begin{equation} \label{inequality 1}
\int_x^{\infty}\frac{u(2+u^2)}{(1+u^2)^{3/2}}e^{-u^2/2} du < \frac{g_1(x)}{g_2(x)}\int_x^{\infty}e^{-u^2/2} du.
\end{equation}
By using Eqns. (\ref{equality 1}) and  (\ref{inequality 1}), we get
\begin{equation}
 \frac{g_1(x)}{g_2(x)}\int_x^{\infty}e^{-u^2/2} du > \frac{1}{\sqrt{1+x^2}}e^{-x^2/2}
\end{equation}
which is equivalent to
\begin{equation}
\int_x^{\infty}e^{-u^2/2} du > \frac{1+x^2}{x(2+x^2)}e^{-x^2/2}
\end{equation}
Thus, the inequality Eqn. (\ref{new low}) is proved.

On the limit case, it is easy to prove
\begin{equation}
\lim_{x\rightarrow\infty} \frac{\frac{1+x^2}{x(2+x^2)}}{\sqrt{1+x^2}}=1
\end{equation}
which indicates that
\begin{equation}
\int_x^{\infty}e^{-u^2/2}du\approx\frac{1}{\sqrt{1+x^2}}e^{-x^2/2}
\end{equation}
is true.

The proof of Theorem 3 is completed.

\section{Tightness Comparison}

It is hard to see that  for $x>1$
\begin{equation}
\frac{1}{x}>\frac{1}{\sqrt{1+x^2}}
\end{equation}
Thus,
\begin{equation}
\frac{1}{x}e^{-x^2/2}>\frac{1}{\sqrt{1+x^2}}e^{-x^2/2}
\end{equation}

This indicates that our new developed inequality in Theorem 3 has a tighter upper bound on the estimation of $\int_x^{\infty}e^{-u^2/2}du$ than that given in Theorem 1.

On the lower bound tightness, it is hard to see that
\begin{equation}
\frac{\frac{x}{1+x^2}}{\frac{1+x^2}{x(2+x^2)}}=\frac{x^2(2+x^2)}{(1+x^2)^2}=\frac{(1+x^2)^2-1}{(1+x^2)^2}<1
\end{equation}
Therefore,
\begin{equation}
\frac{x}{1+x^2} e^{-x^2/2}<\frac{1+x^2}{x(2+x^2)}e^{-x^2/2}
\end{equation}
which means that our new developed inequality in Theorem 3 has a tighter lower bound on the estimation of $\int_x^{\infty}e^{-u^2/2}du$ than that given in Theorem 1.

On the comparison of Theorem 2 and Theorem 3, it is very hard to give a simple proof. One can use numerical analysis to get it.  Therefore, we shall discuss it in Section V by numerical method.

\section{Application to the estimation of inverse Q-function}

Since Q-function is usually used to estimate the error probability, and the error probability is often with value close to zero.  In this part, we mainly focus on the estimation of inverse Q-function for Q-function with very small values.  The estimation problem of the inverse Q-function can be described as follows.

\textit{Inverse Q-function Problem}\\

To find a simple function $f_Q(\alpha)$ with an explicit form so that
\begin{equation}
\mid Q^{-}(\alpha)-f_Q(\alpha)\mid \rightarrow 0
\end{equation}
as $\alpha\rightarrow 0$, where $Q(x)=\alpha$.

By using the definition of the Q-function and the results in Theorem 1, for a very small positive value $\alpha$, we have
\begin{equation}
\frac{1}{x}e^{-x^2/2}\gtrapprox \sqrt{2\pi}\alpha
\end{equation}
where $Q(x)=\alpha$.
It is equivalent to  
\begin{equation} \label{inves ineq1}
\frac{1}{x^2}e^{-x^2}\gtrapprox 2\pi\alpha^2
\end{equation}
Since $p(y)=y\log y$ for $y>$ is monotonically decreasing for $ 0<y<e^{-1}$, 
 we  have
\begin{equation} \label{inequality inverseQ}
\log[\frac{1}{x^2}]\frac{1}{x^2} e^{-x^2} -e^{-x^2} \lessapprox (2\pi\alpha^2)\log{(2\pi\alpha^2)}
\end{equation}
 if
\begin{equation}
\frac{1}{x^2}e^{-x^2}<e^{-1}.
\end{equation}

It has two terms at the left-hand side of Eqn. (\ref{inequality inverseQ}). 
It is not hard to see that when $x$ is very large, the second term will become dominant part.  Thus, one can remove the first term from the left-hand side, we get
\begin{equation}
e^{-x^2}\approx -(2\pi\alpha^2)\log(2\pi\alpha^2)
\end{equation}
which means
\begin{equation}
x\approx \sqrt{-\log(-(2\pi\alpha^2)\log(2\pi\alpha^2))}\geq \sqrt{-\log{(-h(2\pi\alpha^2))}}
\end{equation}

Likewise, by using the upper bound in Theorem 3
\begin{equation}
\int_x^{\infty} e^{-u^2/2}du \approx \frac{1}{\sqrt{1+x^2}}e^{-x^2/2}
\end{equation}
when $x$ is sufficient large, one can also get another approximation of the inverse solution of Q-function by
\begin{equation}\label{upp inverseq}
x\approx \sqrt{-\log{(2\pi\alpha^2(1-\log(2\pi\alpha^2)))}}
\end{equation}

It is worthy to note that by using assumption $\frac{1}{x^2}e^{-x^2}<e^{-1}$, one can get $\frac{1}{x^2}<e^{x^2-1}$. Since $h_1(x)=\frac{1}{x^2}$ is  strictly monotonically decreasing for $x>0$ and $ h_2(x)=e^{x^2-1}$ is strictly monotonically increasing for $x>0$ and $h_1(1)=h_2(1)$. Thus,  the inequality $\frac{1}{x^2}<e^{x^2-1}$ holds is equivalent to that the inequality $x>1$ holds. With this result, one can easily see that the first term in the left-hand side of Eqn. (\ref{inequality inverseQ}), $\log[\frac{1}{x^2}]\frac{1}{x^2}e^{-x^2}<0$. By removing it from Eqn. (\ref{inequality inverseQ}), which may increase the value of the left-hand side in Eqn. (\ref{inequality inverseQ}) and make it close to the value of the right-hand side term in Eqn. (\ref{inequality inverseQ}).

Although it is difficult to give an exact approximation error analysis in theory,  we can use the numerical analysis to observe it. Based on various numerical results, we get the following conclusion, which is expressed as a conjecture (due to less of strict mathematical analysis).\\

\textbf{\textit{Conjecture} 3} {(Inverse Q-function Inequality)}\\

Let $\alpha=Q(x)$ for a positive real number $x$, the inverse solution of the Q-function is given by
$x=Q^{-}(\alpha)$, where $Q^{-}$ represents the inverse function of the Q-function. If $\alpha$ is sufficient small, then
we have
\begin{equation}\label{low1 inverseq}
Q^{-}(\alpha) > \sqrt{-\log{((2\pi\alpha^2)\log(2\pi\alpha^2))}}
\end{equation}
and
\begin{equation}
Q^{-}(\alpha)< \sqrt{-\log{(2\pi\alpha^2(1-\log(2\pi\alpha^2)))}}
\end{equation}
Furthermore, we have
\begin{equation}\label{low2 inverseq}
Q^{-}(\alpha)> \sqrt{-\log{h(2\pi\alpha^2)}}
\end{equation}
where $h(x)$ is  the information entropy function of form $h(x)=-x\log(x)-(1-x)\log(1-x)$ for $0\leq x\leq 1$.

Note that the inequality (\ref{low2 inverseq}) sets up a close relation between the information entropy and the Q-function when integral variable $x$ is very large.

\section{Numerical Results}\label{sec:simulation}

In this section, we shall present some numerical results to check the tightness of our new developed inequalities. Fig. 1 and Fig. 2 present some comparison results by using Theorem 1 and Theorem 3 for $0<x<1.5$ and $x>1.5$, respectively, where \text{Ideal, O-upp, O-low, N-upp, N-low} denote the results obtained by using ideal integral, the upper bound of Theorem 1, the lower bound of Theorem 1, the upper bound of Theorem 3 and the lower bound of Theorem 3, respectively. From Fig. 1, it is easy to see that the upper and lower bounds of Theorem 1 are always true and the lower bound of Theorem 3 is true when $x$ is greater than $\sqrt{2}$ and the upper bound of Theorem 3 is valid when $x$ is greater than 0.7862. These results clearly confirm the validness of Theorem 3. Fig. 2 shows that when $x$ is greater than 1.5, the results of Theorem 3 provides better approximations than that using Theorem 1.  Another observation is that when $x$ is less than 0.7862, using $\frac{1+x^2}{x(2+x^2)} e^{-x^2/2}$ really provides the best approximation to $\int_x^{\infty}e^{-u^2/2}du$ and that when $x$ is greater than 0.7862, using $\frac{1}{\sqrt{1+x^2}} e^{-x^2/2}$ can provide the best approximation to $\int_x^{\infty}e^{-u^2/2}du$ among the four bounds in Theorem 1 and Theorem 3.

Fig. 3 shows some numerical results on the comparison of Theorem 2 and Theorem 3, where all the results are normalized by $\int_x^{\infty} e^{-u^2/2}du$. The legend mark "Integral, BS-upp, BS-low, N-upp, and N-low," denote the results obtained by using ideal integral, the the upper bound of Theorem 3, the lower bound of Theorem 3, the upper bound of Theorem 2 and the lower bound of Theorem 2, respectively. It indicates that the new lower bound in Theorem 3 is tighter than that in Theorem 2, but the upper bound in Theorem 3 is worse than that given in Theorem 2.

Fig. 4 presents some numerical results on the inverse Q-function for $\alpha$ less than $10^{-2}$, where \text{Ideal, UPP, Low1, Low2} denote the results obtained by using ideal inverse Q-function, Eqn. (\ref{upp inverseq}), Eqn.(\ref{low1 inverseq}) and Eqn. (\ref{low2 inverseq}), respectively. From Fig. 4, one can find that  Eqn. (\ref{low1 inverseq}) has the best estimation performance to the inverse Q-function and as the value of $\alpha$  decreases, the three approximates will converge the ideal inverse Q-function rapidly as expected, which confirm our developed theoretical results. Another interesting observation is that Eqn.(\ref{low1 inverseq}) and Eqn.(\ref{low2 inverseq}) provide two lower bounds on the inverse Q-function for $\alpha <10^{-2}$ and Eqn. (\ref{upp inverseq}) provides an upper bound on the inverse Q-function for $\alpha<10^{-2}$. Based on the observation, we expressed it as a Conjecture in Section IV.

\section{Conclusion}\label{sec:conclusions}
In this paper, we have presented two new Mills' ratio inequalities with simple expressions, one lower bound and one upper bound. The new developed lower bound is tighter than that well known results on Mills' ratio obtained by Gordon and Sampford.  As their applications, we also considered the approximation of  inverse solution of the Q-function and presented some useful formulas with simple expressions. Some numerical results confirmed that these approximates can  characterize the property of inverse Q-function very well and provide some upper  and lower bounds when the value of Q-function is relatively small. Finally, we then proposed an conjecture on the inverse solution of Q-function.

\section*{Acknowledgement}
This work was supported by NSFC of China No. 61171064, NSFC of China
No. 61021001 and China Major
State Basic Research Development Program (973 Program) No.
2012CB316100(2).

\bibliographystyle{IEEEtran}


\begin{figure}[!t]
\centering
\includegraphics[width=5.0in]{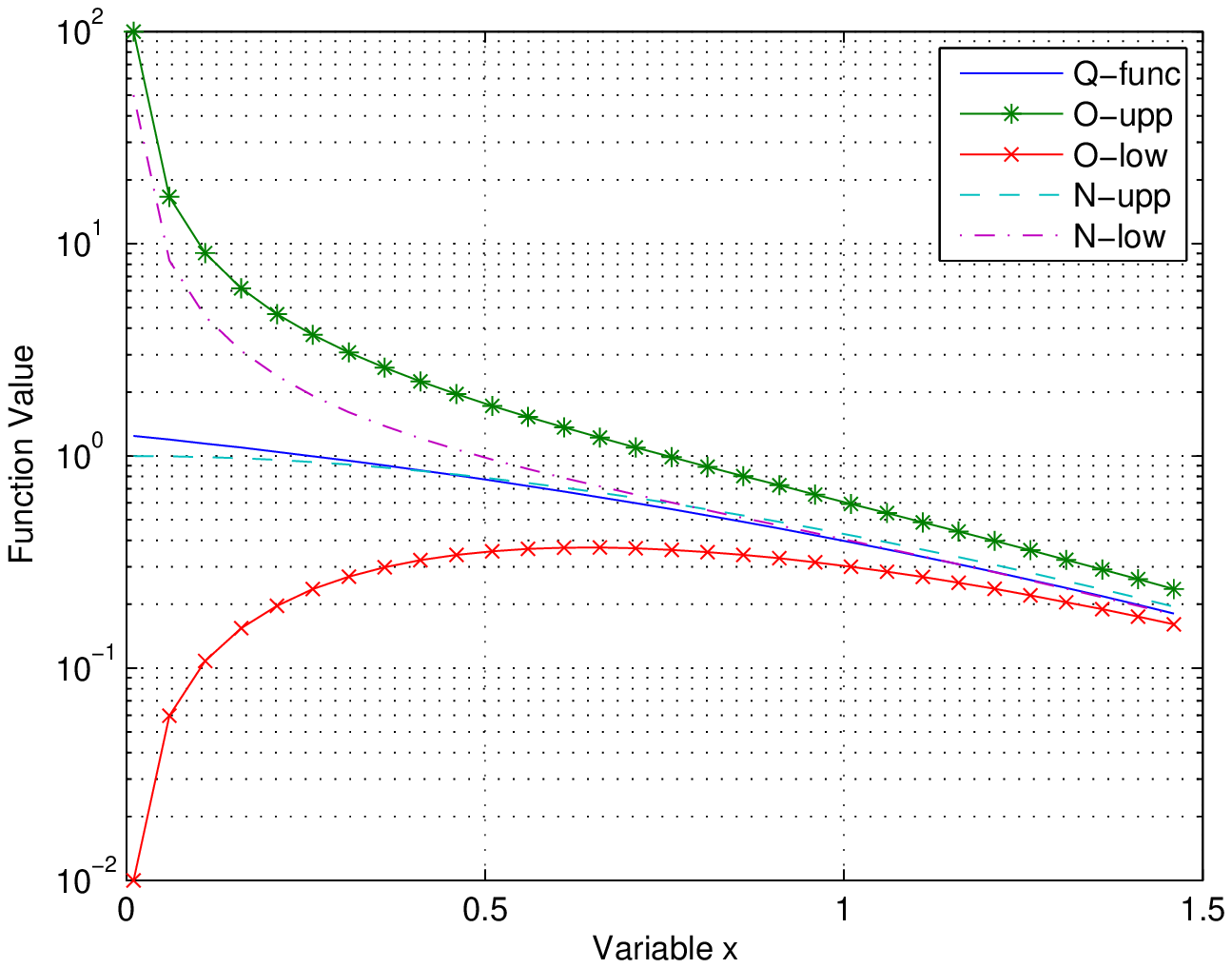}
\caption{Mills' Ratio Approximation for $x$ in the range of $0<x<1.5$} \label{fig:TXm}
\end{figure}

\begin{figure}[!t]
\centering
\includegraphics[width=5.0in]{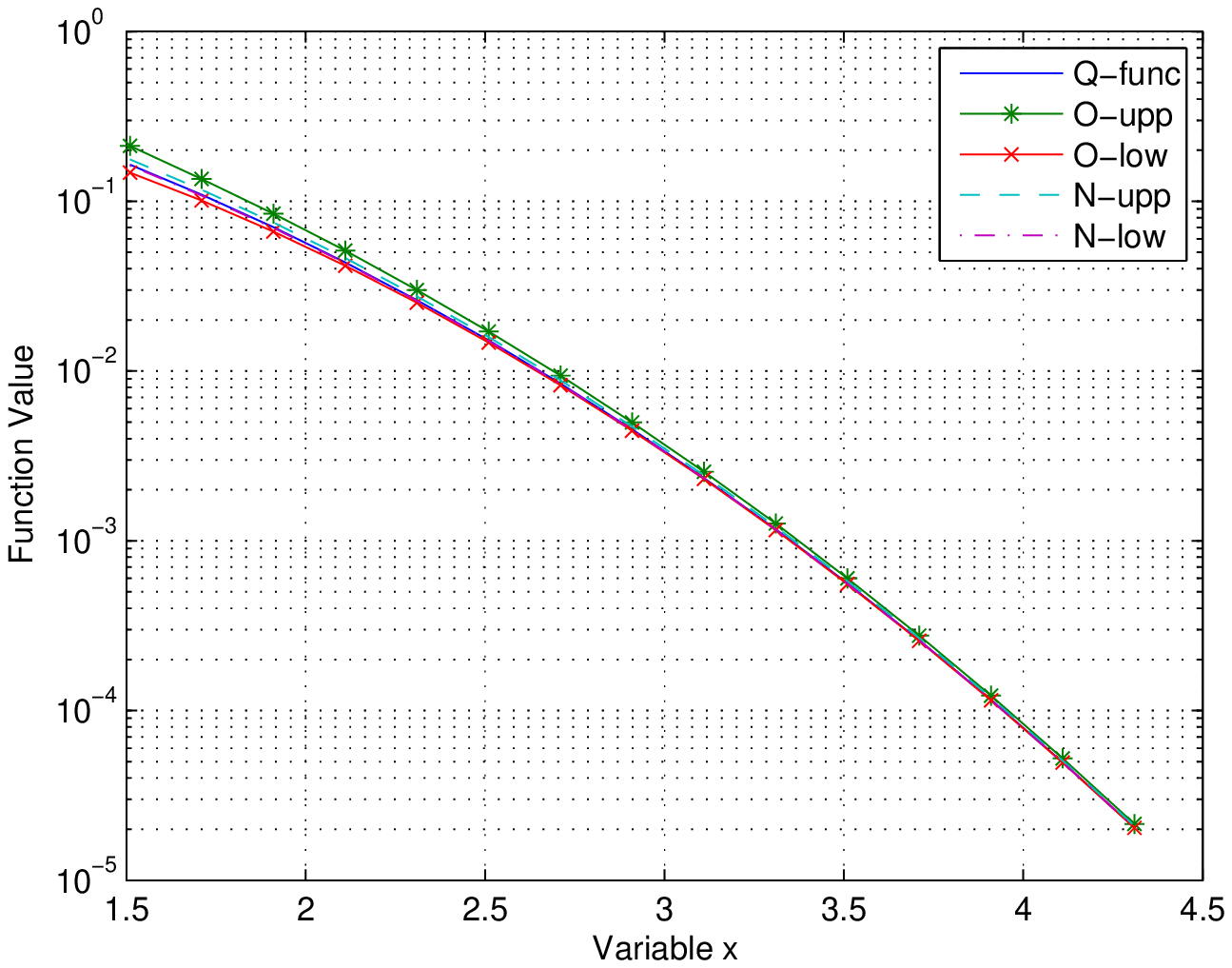}
\caption{Mills' Ratio Approximation for $x$ in the range of $x> 1.5$} \label{fig:TXm}
\end{figure}

\begin{figure}[!t]
\centering
\includegraphics[width=5.0in]{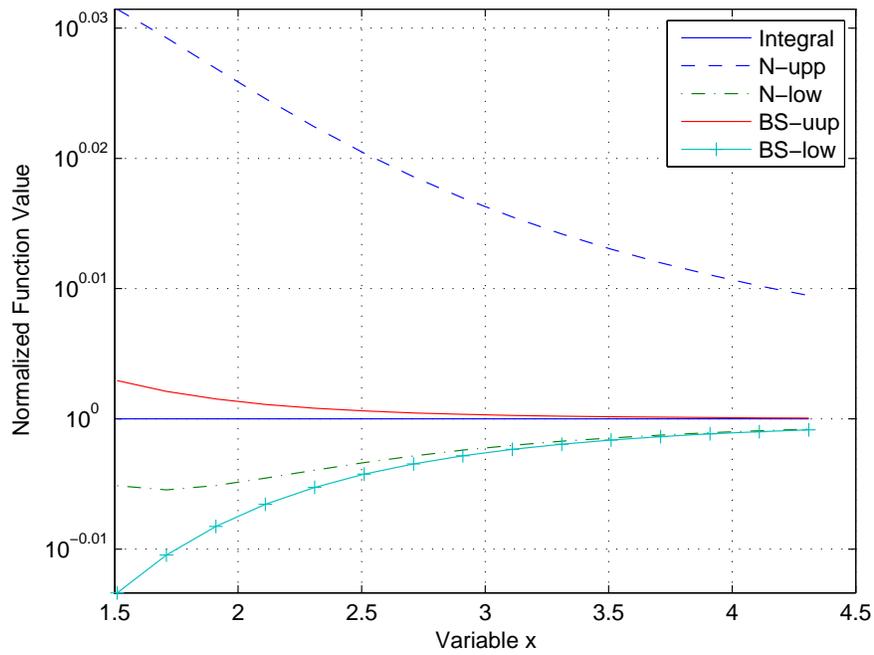}
\caption{Comparison Results on Mills' Ratio Approximation in Theorem 2 and Theorem 3} \label{fig:TXm}
\end{figure}

\begin{figure}[!t]
\centering
\includegraphics[width=5.0in]{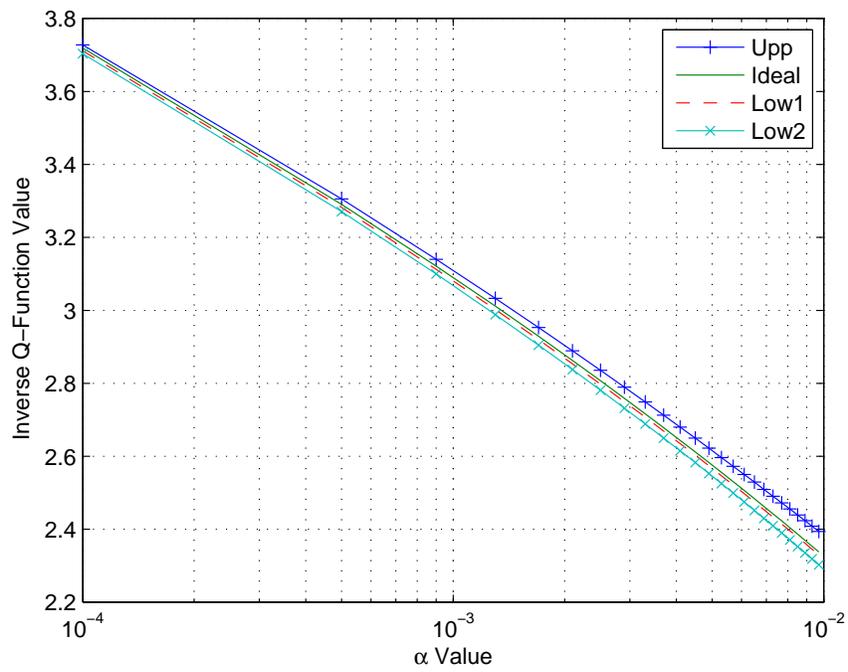}
\caption{Approximation of Inverse Q-Function} \label{fig:TXm}
\end{figure}

\end{document}